\documentstyle[11pt]{article}

\makeatletter
\@addtoreset{equation}{section}
\makeatother

\def\dalemb#1#2{{\vbox{\hrule height .#2pt
        \hbox{\vrule width.#2pt height#1pt \kern#1pt
                \vrule width.#2pt}
        \hrule height.#2pt}}}

\let\a=\alpha    \let\e=\epsilon
  \let\q=\theta  
  \let\n=\nu

\def\nn{\nonumber} \def\bd{\begin{document}} \def\ed{\end{document}}
\def\ds{\documentstyle} \let\fr=\frac \let\bl=\bigl \let\br=\bigr
\let\Br=\Bigr \let\Bl=\Bigl 
\let\bm=\bibitem
\let\na=\nabla
\let\pa=\partial \let\ov=\overline
\def\ie{{\it i.e.\ }} 
\newcommand{\be}{\begin{equation}} 
\newcommand{\ee}{\end{equation}} 
\def\ba{\begin{array}}
\def\ea{\end{array}}
\def\ft#1#2{{\textstyle{{\scriptstyle #1}\over {\scriptstyle #2}}}}
\def\fft#1#2{{#1 \over #2}}
\def\del{\partial}
\def\sst#1{{\scriptscriptstyle #1}}
\def\oneone{\rlap 1\mkern4mu{\rm l}}
\def\e7{E_{7(+7)}}
\def\td{\tilde}
\def\wtd{\widetilde}
\def\im{{\rm i}}
\def\bog{Bogomol'nyi\ }
\def\q{{\tilde q}}
\def\hast{{\hat\ast}}
\def\0{{\sst{(0)}}}
\def\1{{\sst{(1)}}}
\def\2{{\sst{(2)}}}
\def\3{{\sst{(3)}}}
\def\4{{\sst{(4)}}}
\def\5{{\sst{(5)}}}
\def\6{{\sst{(6)}}}
\def\7{{\sst{(7)}}}
\def\8{{\sst{(8)}}}
\def\n{{\sst{(n)}}}

\newcommand{\ho}[1]{$\, ^{#1}$}
\newcommand{\hoch}[1]{$\, ^{#1}$}
\newcommand{\bea}{\begin{eqnarray}} 
\newcommand{\eea}{\end{eqnarray}} 
\newcommand{\ra}{\rightarrow}
\newcommand{\lra}{\longrightarrow}
\newcommand{\Lra}{\Leftrightarrow}
\newcommand{\ap}{\alpha^\prime}
\newcommand{\bp}{\tilde \beta^\prime}
\newcommand{\tr}{{\rm tr} }
\newcommand{\Tr}{{\rm Tr} } 
\newcommand{\NP}{Nucl. Phys. }
\newcommand{\tamphys}{\it Center for Theoretical Physics,
Texas A\&M University, College Station, Texas 77843}
\newcommand{\ens}{\it Laboratoire de Physique Th\'eorique de l'\'Ecole
Normale Sup\'erieure\hoch{3}\\
24 Rue Lhomond - 75231 Paris CEDEX 05}

\newcommand{\auth}{M.J. Duff\hoch{\ddagger1}, H. L\"u\hoch{\dagger} and 
C.N. Pope\hoch{\ddagger2}}

\begin{document}
\begin{flushright}
\hfill{CTP TAMU-09/98}\\
\hfill{LPTENS-98/03}\\
\hfill{hep-th/9803061}\\
\hfill{March 1998}\\
\end{flushright}

\vspace{20pt}

\begin{center}
{\large {\bf AdS$_5\times S^5$ Untwisted}} 

\vspace{30pt}

\auth

\vspace{15pt}
{\hoch{\dagger}\ens}

\vspace{10pt}
{\hoch{\ddagger}\tamphys}

\vspace{40pt}

\underline{ABSTRACT}
\end{center}

      Noting that $T$-duality untwists $S^{5}$ to $CP^{2} \times
S^{1}$, we construct the duality chain: $n=4$ super Yang-Mills
$\rightarrow$ Type IIB superstring on AdS$_{5} \times S^{5}$
$\rightarrow $ Type IIA superstring on AdS$_{5} \times CP^{2} \times
S^{1}$ $\rightarrow $ M-theory on AdS$_{5} \times CP^{2} \times
T^{2}$. This provides another example of {\it supersymmetry without
supersymmetry}: on AdS$_{5} \times CP^{2} \times S^{1}$, Type IIA {\it
supergravity} has $SU(3) \times U(1)\times U(1) \times U(1)$ and $N=0$
supersymmetry but Type IIA {\it string theory} has $SO(6)$ and
$N=8$. The missing superpartners are provided by stringy winding
modes.  We also discuss IIB compactifications to AdS$_{5}$ with
$N=4$, $N=2$ and $N=0$.

{\vfill\leftline{}\vfill
\vskip	10pt
\footnoterule
{\footnotesize
        \hoch{1} Research supported in part by NSF Grant PHY-9722090.
\vskip  -12pt} \vskip   14pt
{\footnotesize
	\hoch{2}	Research supported in part by DOE 
grant DE-FG03-95ER40917. \vskip	-12pt}  \vskip	14pt
{\footnotesize
        \hoch{3} Unit\'e Propre du Centre National de la Recherche
Scientifique, associ\'ee \`a l'\'Ecole Normale Sup\'erieure
\vskip -12pt} \vskip 10pt
{\footnotesize \hoch{\phantom{3}} et \`a l'Universit\'e de Paris-Sud.
\vskip -12pt} \vskip 10pt}

\pagebreak
\setcounter{page}{1}

\section{Introduction}

     There is now persuasive evidence that certain conformal field
theories in various dimensions are dual to M/string theory
compactified to anti-de Sitter space \cite{Maldacena}. In particular
$(n=4,d=4)$ super Yang-Mills theory is dual to type IIB string theory
on AdS$_{5} \times S^{5}$.  In this paper we note that $S^{5}$ may be
regarded as a $U(1)$ bundle over $CP^{2}$ and hence that the theory
admits a $T$-dual type IIA description.  In fact $T$-duality untwists
$S^{5}$ to $CP^{2} \times S^{1}$. Thus we can construct the duality
chain: $n=4$ super Yang-Mills $\rightarrow$ type IIB superstring on
AdS$_{5} \times S^{5}$ $\rightarrow$ type IIA superstring on AdS$_{5}
\times CP^{2}\times S^{1}$ $\rightarrow $ $M$-theory on AdS$_{5}
\times CP^{2}\times T^{2}$.  This provides another example of the 
phenomenon of {\it supersymmetry without
supersymmetry} \cite{DLP}, but this time without involving Dirichlet
0-branes.   On AdS$_{5} \times CP^{2} \times S^{1}$
type IIA {\it supergravity} has $SU(3) \times U(1)\times U(1)\times
U(1)$ and $N=0$
supersymmetry. Indeed, since $CP^{2}$ does not admit a spin structure,
its spectrum contains no fermions at all! Nevertheless, type IIA {\it
string theory} has $SO(6)$ and $N=8$ supersymmetry. The missing
superpartners (and indeed all the fermions) are provided by stringy
winding modes.  These winding modes also enhance $SU(3)\times U(1)$ to
$SO(6)$, while the gauge bosons of the remaining $U(1)\times U(1)$
belong to massive multiplets.

     We also describe the way in which $p$-brane solutions of type IIA
and type IIB are related when the compactification takes the form of a
$U(1)$ fibration rather than an $S^{1}$, which is the case, for
example, for all odd-dimensional spheres.

     In addition to the $S^{5}$ compactification of type IIB, which
has the maximal $N=8$ supersymmetry in $D=5$, we also exhibit new
compactifications on the spaces $Q(n_1,n_2)$, which are $U(1)$ bundles over
$S^{2} \times S^{2}$, with winding numbers $n_1$ and $n_2$ respectively.
Generically, these have $N=0$ supersymmetry, but the special
case $Q(1,1)$ has $N=4$ supersymmetry.  Reversing the
orientation of $Q(1,1)$, we get a solution with $N=0$ supersymmetry.
However, if these are dual to $(n=2,d=4)$ or $(n=0,d=4)$ 
super Yang-Mills theories with vanishing $\beta$ function, they would have
to be unusual ones with non-integer conformal dimensions.  We also
present compactifications that give $N=2$ and $N=0$  supersymmetries, but with
integer conformal dimensions, in which $S^5$ is replaced by a lens space.
(Reducing the supersymmetries by using lens spaces was discussed in
\cite{DNP3}.) 

     The organisation of this paper is as follows.  Since
supergravity and supermembranes on AdS and their relation to
singletons and superconformal theories was an active area of research
some years ago, we begin with a brief review in section 2.  Then, in
section 3, we introduce some ideas and notation that will be relevant
also in the later parts of the paper.  Specifically, we consider a
third kind of dimensional reduction for $p$-branes, in addition to the
usual double-dimensional reduction and vertical dimensional reduction
schemes, which is applicable to cases where the transverse space is
even dimensional.  Viewing the transverse space as a foliation of
spheres, we note that in these cases the spheres are odd dimensional,
and hence can be viewed as $U(1)$ bundles over complex projective
spaces.  We can perform a dimensional reduction on the $U(1)$
direction; we shall refer to this as ``Hopf reduction.''  Combined
with a T-duality transformation, this reduction allows the
``untwisting'' of the $U(1)$ fibres.  In section 4, we consider the
near-horizon limit of this kind of reduction, in the case of the
self-dual 3-brane of the type IIB theory \cite{HS,DLgauge}.  This corresponds
 to the
AdS$_5\times S^5$ solution, and we show how it may be mapped by
T-duality into an AdS$_5\times CP^2\times S^1$ solution of the type
IIA theory, and hence to an AdS$_5\times CP^2\times T^2$ solution in
M-theory.  We then discuss the spectrum of states in the dual
description, and show in particular how the fermions, including the
gravitini, are present only in the string theory or M-theory spectrum,
but not in the supergravity spectrum.  Many other solutions of the
form AdS$_5\times M_5$ exist, where $M_5$ is any Einstein space (which
need not even be homogeneous).  In section 5 we consider an infinite
family of examples, where $M_5$ is the space $Q(n_1,n_2)$ described by
the $U(1)$ bundle over $S^2\times S^2$ that has winding numbers $n_1$
and $n_2$ over the two 2-sphere factors in the base.  We show that the
space $Q(1,1)$ gives an $N=4$ supersymmetric solution, and that with an 
orientation reversal, it gives $N=0$.  These type IIB
solutions can be mapped via T-duality to type IIA solutions of the
form AdS$_5\times S^2\times S^2 \times S^1$.  We also obtain solutions on
the products of AdS$_5$ and lens spaces, with $N=2$ and $N=0$.
In section 6, we give
further examples of solutions of the form AdS$_5\times$spheres,
including some that arise as the near-horizon limit of
multiply-charged intersecting $p$-brane solutions.  Finally, after our
conclusions, we give the detailed form of the T-duality mapping
between the type IIA and type IIB theories in an appendix.

\section{Anti-de-Sitter space, branes, singletons, superconformal field
theories and all that}

In the early 80's there was great interest in $N$-extended
supergravities for which the global $SO(N)$ is promoted to a gauge
symmetry \cite{dasfree}, in particular the maximal $N=8$, $SO(8)$ theory
\cite{DN}. In these theories the underlying symmetry is described by
the $D=4$ anti-de Sitter (AdS$_4$) supersymmetry algebra, and the
Lagrangian has a non-vanishing cosmological constant proportional to
the square of the gauge coupling constant.  This suggested that there
might be a Kaluza-Klein interpretation, and indeed this
maximal theory was seen to correspond to the massless sector of $D=11$
supergravity compactified on an $S^7$ whose metric admits an $SO(8)$
isometry \cite{DP}. An important ingredient in these developments that
had been insufficiently emphasized in earlier work on Kaluza-Klein
theory was that the AdS$_4 \times S^7$ geometry was not fed in by hand
but resulted from a {\it spontaneous compactification}, i.e. the
vacuum state was obtained by finding a stable solution of the
higher-dimensional field equations \cite{CS}.  The mechanism of
spontaneous compactification appropriate to the AdS$_4 \times S^7$
solution eleven-dimensional supergravity was provided by the 
Freund-Rubin mechanism \cite{FR} in which the $4$-form field strength in
spacetime $F_{\mu\nu\rho\sigma}$ ($\mu=0,1,2,3$) is proportional to
the alternating symbol $\epsilon_{\mu\nu\rho\sigma}$ \cite{DV}.  By
applying a similar mechanism to the $7$-form dual of this field
strength one could also find compactifications on AdS$_{7} \times
S^{4}$ \cite{PTV} whose massless sector describes gauged maximal
$N=4$, $SO(5)$ supergravity in $D=7$ \cite{PPV,TV}. A summary of these
AdS compactifications of $D=11$ supergravity may be found in
\cite{DNP}. Type IIB supergravity in $D=10$, with its self-dual
$5$-form field strength, also admits a Freund-Rubin compactification
on AdS$_{5}\times S^{5}$ \cite{GM,KRV} whose massless sector describes
gauged maximal $N=8$ supergravity in $D=5$ \cite{PPV2,GRW}.

In the three cases given above, the symmetry of the vacuum is
described by the supergroups $OSp(4/8)$, $SU(2,2/4)$ and $OSp(6,2/4)$
for the $S^7$, $S^5$ and $S^4$ compactifications respectively. Each of
these groups is known to admit the so-called singleton or doubleton
representations \cite{Fronsdal,gun}.  Curiously, although they appeared in
the Kaluza-Klein harmonic expansions \cite{Sezgin,GRW2}, they could be
gauged away. In fact they reside not in the bulk of AdS but on the
boundary \cite{Fronsdal} where the above supergroups correspond to the
superconformal groups \cite{Nahm}. In the case of $S^7$, one finds an
$(n=8,d=3)$ supermultiplet with $8$ scalars and $8$ spinors; in the
case of $S^5$ one finds a $(n=4,d=4)$ supermultiplet with $1$ vector,
$4$ spinors and $6$ scalars, and in the case of $S^4$ one finds a
$((n_+,n_-)=(2,0),d=6)$ supermultiplet with $1$ $2$-form with
self-dual field strength, $8$ spinors and $5$ scalars.

With the discovery of the eleven-dimensional supermembrane \cite{BST},
it was noted that the physical degrees of freedom on the worldvolume
of the membrane also correspond to the $(n=8,d=3)$ supermultiplet with
$8$ scalars and $8$ spinors and it was conjectured that these may in
fact admit the interpretation of singletons \cite{Fifteen}. There
followed a good deal of activity relating super $p$-branes on
AdS$_{p+2} \times S^{D-p-2}$,
singletons and superconformal field theories
\cite{BDPS,BDPS2,NS,BD,Classical,BD2,BSTan,BSS,NST,DPS}.  In particular,
it was possible to find solutions of the combined supergravity
supermembrane equations describing a membrane occupying the $S^{1}
\times S^{2}$ boundary of AdS$_{4}$: the ``membrane at the end of the
universe'' \cite{BDPS,BDPS2}. The action and transformation rules for
the $OSp(4/8)$ singleton conformal field theory were presented in
\cite{BD} and a general correspondence between super $p$-branes and
the superconformal field theories in Nahm's classification \cite{Nahm}
was discussed in \cite{BD,NST,BD2}.

These early works focussed on scalar supermultiplets because these
were the only $p$-branes known at the time \cite{AETW}. However, with
the discovery of type $II$ $p$-brane solitons
\cite{CHS1,CHS2,HS,DLgauge,Luscan}, vector
and tensor multiplets were also seen to play a role. In particular, the
worldvolume fields of the self-dual type IIB superthreebrane were shown to
be described by a $(n=4,d=4)$ gauge theory \cite{DLgauge}. The
subsequent realisation that this theory admitted the interpretation of
a Dirichlet $3$-brane \cite{P}, and the observation that the
superposition of $N$ such branes yields an $SU(N)$ gauge theory
\cite{witt} are, of course, crucial to the duality with type IIB theory
\cite{Maldacena}. For earlier related work on coincident threebranes 
and $n=4$ super Yang Mills, see 
\cite{Gubser1,Klebanov,Gubser2,Gubser3}

More recently, AdS has emerged in the near-horizon geometry of black
$p$-brane solutions \cite{GT,DGT,GHT,DKL} in $D$ dimensions. The dual
brane, with worldvolume dimension ${\tilde p}+1=D -p-3$, interpolates
between $D$-dimensional Minkowski space and AdS$_{{\tilde p}+2}\times
S^{p+2}$ (or $M_{{\tilde p}+2}\times S^{3}$ if $p=1$). It is the
subset of those solutions with constant or zero dilaton, the
non-dilatonic $p$-branes, that have been conjectured to be dual to
conformal theories with vanishing beta function
\cite{Maldacena}. Generically, however, the gradient of the dilaton
plays the role of a conformal Killing vector on AdS \cite{DGT} and
these may be related to theories with non-vanishing beta function
\cite{IMSY}. Moreover, new super $p$-branes with fewer
supersymmetries may also be constructed which interpolate between
$D$-dimensional flat space and AdS$ \times M$ where $M$ is any
Einstein space \cite{DLPS,ccdfft}, not necessarily a round sphere. For
example $N=1$ for the squashed $S^{7}$ \cite{ADP,DNP2}.  Note that the
space at large distance, although asymptotically locally flat, is not
asymptotic to Minkowski spacetime.  Rather, it approaches a flat
metric on a generalised cone \cite{DLPS}.

Since in gauged supergravity the gauge coupling is related to the AdS
cosmological constant, the beta function is determined by the
renormalisation of the cosmological constant \cite{CD} which in turn
is fixed by the Weyl anomaly \cite{Weyl,DDI}. The Weyl anomaly at
one-loop (and indeed all odd-loop orders) vanishes trivially in the
case of $(N=4, D=7)$ and $(N=8,D=5)$ supergravities, since the
spacetime has odd dimension \cite{Weyl}.  One might naively have
expected to find a non-vanishing beta function for gauged $(N=8,D=4)$
supergravity, but remarkably it vanishes \cite{CDGR}.  This result
continues to hold when the massive Kaluza-Klein multiplets are
included \cite{GN,IY}, by virtue of the $N>4$ spin-moment sum rules
\cite{Curtwright,Supergravity81}.  Moreover, the argument has been
extended to all orders \cite{ST}.

This vanishing of the beta function in the AdS$_{4} \times S^{7}$
compactification is an answer that has been looking for a question for
eighteen years \cite{CDGR}. Now, however, we see that this is entirely
consistent with the recent conjectured duality between supergravity on
AdS and certain superconformal field theories whose coupling constant
in given by the AdS cosmological constant \cite{Maldacena}, a duality
that forms the subject of the present paper.

Following Maldacena's conjecture \cite{Maldacena}, a number of papers
appeared reviving the old singleton-AdS-membrane-superconformal field
theory connections
\cite{Ferrara2,KKR,Boonstra,CKKTV,IMSY,Gunaydin,Gubser,%
Horowitz,Witten,Kachru,Berkooz,Ferrara,Rey,Maldacena2,lnv} and applying
them to this new duality context.
In particular, the there is seen to be a correspondence between the
Kaluza-Klein mass spectrum in the bulk and the conformal dimension of
operators on the boundary \cite{Gubser,Witten}. The philosophy is that 
supergravity is a
good approximation for large $N$ and that stringy excitations
correspond to operators whose dimensions diverge for $N \rightarrow
\infty $. Under the IIB/IIA $T$-duality discussed in the present paper,
however, the Kaluza-Klein and certain stringy excitations trade places and so
the type IIA (or $D=11$) supergravity picture may throw some light on the
finite $N$ regime.

\section{Hopf fibrations of M-theory and string theory}

     There are two kinds of dimensional reduction of $p$-brane
solitons that are commonly considered.  The simpler is double dimensional
reduction, where a Kaluza-Klein reduction on a $p$-brane
worldvolume coordinate is performed.  This is always possible, since
the $p$-brane solitons have translational isometries on their
worldvolumes.  The effect is to reduce a $p$-brane in $(D+1)$ dimensions
to a $(p-1)$-brane in $D$ dimensions.  The second kind of
dimensional reduction, known as vertical reduction, involves
performing the Kaluza-Klein reduction in a transverse-space
direction instead.  In order to do this, it is necessary first to
construct a $p$-brane solution in $(D+1)$-dimensions with an isometry
along a transverse-space direction.  This can be done by considering a
multi-soliton configuration, with a uniform continuum of $p$-branes
along the chosen reduction axis.  In this reduction scheme, a
$p$-brane in $(D+1)$ dimensions is reduced to a $p$-brane in $D$
dimensions.

    In this section, we shall consider various examples of a third
kind of dimensional reduction scheme, which is possible whenever the
transverse space has dimension $2n$ that is even.  This means that in
hyperspherical polar coordinates, the usual flat metric $dy^i\, dy^i$
can be written in the form $dr^2 + r^2\, d\Omega_{2n-1}^2$, where 
$d\Omega_{2n-1}^2$ is the metric on the unit $(2n-1)$-sphere.  Now the
odd-dimensional sphere $S^{2n-1}$ can be written as a $U(1)$ bundle
over $CP^{n-1}$, and we may then perform a Kaluza-Klein reduction on
the $U(1)$ direction of the Hopf fibres, since this corresponds to an 
isometry of the sphere, and hence of the transverse space.  This
``Hopf'' reduction scheme is
more akin to the vertical reduction described above, in that it
involves a reduction in the transverse space, while the Poincar\'e 
symmetry on the $p$-brane worldvolume survives unscathed.  However, 
unlike normal
vertical reduction, it is not necessary first to construct a uniform
distribution of $p$-branes; a single $p$-brane solution in $D$
dimensions already has the necessary $U(1)$ isometry.

     In the following subsections, we shall consider examples
where we perform Hopf reductions on $p$-branes in M-theory, the type
IIA string, and the type IIB string.

\subsection{Hopf reductions in M-theory}

    We shall take as our starting point the BPS-saturated membrane
solution of $D=11$ supergravity.  The bosonic Lagrangian of
11-dimensional supergravity contains the metric and a 4-form field
strength $F_4=dA_3$, and is given by \cite{cjs}
\be
{\cal L} = e R - \ft1{48} F_4^2 + \ft1{(12)^4}\, \epsilon^{{\sst M}_1\cdots 
{\sst M}_{11}}\, F_{{\sst M}_1\cdots {\sst M}_4}\, F_{{\sst M}_5\cdots 
{\sst M}_8}\, A_{{\sst M}_9\cdots {\sst M}_{11}}\ .
\label{11lag}
\ee
It admits the extremal membrane solution \cite{dust}
\bea
ds_{11}^2 &=& H^{-2/3}\, dx^\mu dx^\nu \eta_{\mu\nu} + H^{1/3}\, (dr^2 + r^2
d\Omega_7^2 )\nonumber\\
F_4 &=& d^3x\wedge dH^{-1} = Q\, \ast\Omega_7
\ ,\label{membrane}
\eea
where $H=c + \ft16 Q\, r^{-6}$ is harmonic in the 8-dimensional
transverse space whose flat metric is described in terms of the radial
coordinate $r$ and the metric $d\Omega_7^2$ on the unit 7-sphere.  We
are using $\Omega_7$ to denote the volume form on the unit 7-sphere,
$\ast$ denotes the Hodge dual, and $Q$ is the electric charge carried
by the membrane. Conventionally, the constant $c$ is chosen to be
unity, so that the metric tends to the standard eleven-dimensional
Minkowski metric as $r$ tends to infinity.

    The odd-dimensional spheres can be viewed as Hopf fibrations over
complex projective spaces, and specifically, $S^7$ is a $U(1)$ bundle
over $CP^3$. Indeed the standard unit-radius metric $d\Omega_7^2$ on
$S^7$ can be written as:
\be
d\Omega_7^2 = d\Sigma_6^2 + (dz + \bar{\cal A})^2\ ,\label{hopf7}
\ee
where $d\Sigma_6^2$ is the standard Fubini-Study metric on $CP^3$, and
the Kaluza-Klein vector potential $\bar{\cal A}$ has field strength
$\bar{\cal F}$ given by $\bar{\cal F}=2J$, where $J$ is the K\"ahler
form on $CP^3$.  The Fubini-Study metric $\bar g_{ij}$ on $CP^3$ is
Einstein, and we choose a normalisation where its Ricci tensor
satisfies $\bar R_{ij}= 8 \bar g_{ij}$.  The potential is given in
terms of its components by $\bar{\cal A} = \bar{\cal A}_{i} dy^{i}$,
where $y^{i}$ are the coordinates on $CP^3$ and the K\"ahler form
satisfies $\bar\nabla_i J_{jk}=0$ and $J_i{}^j\, J_j{}^k
=-\delta_i^k$.  The coordinate $z$ has period $4\pi$.

    Using (\ref{hopf7}), the membrane metric (\ref{membrane}) can be
written as
\be
ds_{11}^2 = H^{-2/3}\, dx^{\mu} dx^\nu \eta_{\mu\nu} +
            H^{1/3} (dr^2 + r^2 d\Sigma_6^2) + 
            H^{1/3}\, r^2 (dz+ \bar{\cal A})^2\ .\label{mem2}
\ee
We may now compactify the solution on the circle parameterised by the
$U(1)$ fibre coordinate $z$.  In general, the Kaluza-Klein reduction
of a metric from $D+1$ to $D$ dimensions takes the form $ds_{\sst
D+1}^2 = e^{-2\a\varphi}\, ds_{\sst D}^2 + e^{2(D-2)\a\varphi}\,
(dz+\bar{\cal A})^2$, where $\a=((2(D-1)(D-2))^{-1/2}$, and the
parameterisation is such that the lower-dimensional metric is in the
Einstein frame (\ie $\sqrt{g_{\sst D+1}}\, R_{\sst D+1}$ reduces to
$\sqrt{g_{\sst D}} \, R_{\sst D}$).  Applying this to (\ref{mem2}), we
see that the dimensionally-reduced ten-dimensional solution is given
by
\bea
ds_{10}^2 &=& 
H^{-5/8}\, r^{\ft14}\, dx^\mu dx^\nu \eta_{\mu\nu} +
H^{3/8}\, r^{\ft14}(dr^2 + r^2 d\Sigma_6^2)\ ,\nonumber\\
e^{4\varphi} &=& Hr^6\ ,\qquad
F_4 = d^3x\wedge d H^{-1} = Q\, e^{-\ft12\varphi} \, \ast\Sigma_6\ ,\qquad
{\cal F} = 2 J\ ,\label{mem1d10}
\eea
where $\Sigma_6$ is the volume form of $CP^3$.  Note from
(\ref{hopf7}) that it is related to the volume form of the 7-sphere by
$\Omega_7=(dz+\bar{\cal A})\wedge\Sigma_6 =dz\wedge \Sigma_6$.

     Unlike the usual membrane solution in $D=10$ type IIA string,
which could be obtained by vertical dimensional reduction of the
membrane in $D=11$, the field strength ${\cal F}$ of the Kaluza-Klein
vector ${\cal A}$, associated with the K\"ahler form on the $CP^3$,
also acquires a charge, in addition to the charge carried by the
4-form field strength.  Nevertheless, since the $U(1)$
compactification of M-theory gives rise to the IIA string, it follows
that the above configuration solves the equations of motion of IIA
massless supergravity in $D=10$ \cite{nilpop}.  Up to the conformal
factor $r^{1/4}$, its metric is reminiscent of the usual membrane
solution in $D=10$, except for the fact that it is the $CP^3$ metric
$d\Sigma_6^2$, rather than the unit 6-sphere metric $d\Omega_6^2$,
that appears in the part describing the transverse space.  Although 
it came by dimensional reduction from a BPS solution of $D=11$ 
supergravity preserving 16 of the 32 components of supersymmetry, it
does not itself 
preserve 16 as a solution of type IIA {\it supergravity}. This is because  
some of the Killing spinors are described by Dirichlet $0$-branes that 
are absent in the type IIA supergravity picture, and indeed they are
absent in perturbative type IIA string theory.  In fact the solution will 
preserve either 12 or 0 components of supersymmetry, depending on the 
orientation of the $CP^{3}$ 
\cite{DLP}. Of course, all 16 will be present in the non-perturbative type IIA 
string theory.  The moral is that neither type IIA supergravity nor 
perturbative type IIA string theory is always a 
reliable guide to the number of supersymmetries preserved in M-theory.

          The ten-dimensional solution (\ref{mem1d10}) can be further 
compactified on $CP^3$, giving rise to the four-dimensional metric 
\be
ds_4^2 = H^{-1/2}\, r\, dx^\mu dx^\nu \eta_{\mu\nu} +
H^{1/2}\, r\, dr^2\ .
\ee

   Let us now return for a moment to the eleven-dimensional membrane
solution (\ref{membrane}).  There is an horizon at $r=0$, and in this
neighbourhood the metric $ds_{11}^2$ is of the form AdS$_4\times S^7$,
where AdS$_4$ is four-dimensional anti-de Sitter spacetime.  This may
be seen by noting that at sufficiently small $r$ the constant $c$ in
the harmonic function $H=c +\ft16 Q\, r^{-6}$ becomes negligible, and
indeed if we temporarily set $c=0$ the metric $ds_{11}^2$ becomes
\be
ds_{11}^2 = k^{\ft13} \, \Big( \ft1k e^{4\rho} \, dx^\mu dx^\nu
\eta_{\mu\nu} + d\rho^2 \Big) + k^{\ft13}\, d\Omega_7^2\ ,\label{ads}
\ee
where $\rho=\log r$ and $k=\ft16 Q$.  The terms inside the first bracket
describe the  metric on AdS$_4$ written in horospherical \cite{DGT,GHT,lpt} 
coordinates $x^\mu$ and $\rho$.    Thus the membrane solution
(\ref{membrane}) with $c=1$ can be viewed as interpolating between
AdS$_4\times S^7$ at the horizon and eleven-dimensional Minkowski spacetime
at radial infinity \cite{GT,DGT,GHT}.

     The AdS$_4\times S^7$ solution (\ref{ads}) can be reinterpreted
as an AdS$_4$ solution of gauged $N=8$ supergravity in four
dimensions, since this latter theory is obtained by Kaluza-Klein
reduction of eleven-dimensional supergravity on the 7-sphere
\cite{DP,DNP}.  On the other hand, it can also be viewed first of all as
an AdS$_4\times CP^3$ solution of the type IIA string, and then, by
compactifying on the $CP^3$, we again obtain a theory on AdS$_4$ in
four dimensions.  This can be seen from the discussion in this
section, by considering the near-horizon structure of the
dimensionally-reduced solution (\ref{mem1d10}).  It was first studied
in the supergravity context in \cite{nilpop}, and more recently an
M-theory discussion was given in \cite{DLP}.

    This completes our discussion of the Hopf reduction of the
membrane solution in M-theory.  Note that we cannot carry out an
analogous Hopf reduction of the 5-brane, since now the transverse
space has dimension 5, and is a foliation of 4-spheres, which cannot
be described as Hopf bundles.

\subsection{Hopf reductions of type II $p$-branes}

     We now turn to a consideration of the $p$-branes of the type IIA and 
type IIB strings in ten-dimensions, and show how in certain cases they may 
be viewed as $U(1)$ fibrations from a nine-dimensional point of view.  In 
these cases we may establish a correspondence between a $p$-brane solution 
of the type IIA or type IIB theory and a different kind of solution of the 
type IIB or type IIA theory, respectively, by making use of the T-duality 
between the type IIA and type IIB strings compactified on a 
circle.\footnote{The idea of generating new solutions from old by
means of T-duality
transformations has been considered in various contexts.  See, for
example, \cite{aal,bho}.}
Specifically, these correspondences may be implemented whenever the
transverse space in the $p$-brane solution is even-dimensional, implying
that it is described in hyperspherical coordinates in terms of a foliation
of odd-dimensional spheres.  Then, in a manner analogous to the 7-sphere
discussion in the previous section, we can compactify on the $U(1)$ fibre
coordinate of the sphere. 

     The $p$-brane solitons in the type IIA theory arise for 
$p=\{0_{\rm\sst D}, 1, 2_{\rm\sst D}, 4_{\rm\sst D},5, 6_{\rm\sst D},
8_{\rm\sst D}\}$, corresponding to foliations of the spheres $\{S^8, S^7, 
S^6, S^4, S^3, S^2, S^0\}$ respectively.  (The subscripts in the list of 
$p$-branes indicate the ones that are D-branes.  The D8-brane arises only in 
the massive IIA supergravity.)  For the type IIB string, the $p$-branes are
$\{-1_{\rm\sst D}, 1, 1_{\rm\sst D}, 3_{\rm\sst D}, 5, 5_{\rm\sst D}, 
7_{\rm\sst D}\}$, associated with the spheres $\{S^9, S^7, S^7, S^5, S^3, 
S^3, S^1\}$ respectively.  Note that in the type IIB case all the $p$-branes 
are associated with odd-dimensional spheres, and so they may all be 
compactified on a $U(1)$ fibre coordinate.  On the other hand in the type 
IIA case only the string and the 5-brane are associated with odd-dimensional 
spheres, whilst all the type IIA D-branes are associated with
even-dimensional spheres.  

     As a preliminary, we shall show how to construct the odd-dimensional 
unit spheres $S^{2n+1}$ as $U(1)$ bundles over $CP^n$.  The construction, 
which generalises the 7-sphere example that we used in the previous section, 
involves writing the metric $d\Omega_{2n+1}^2$ on the unit $(2n+1)$-sphere
in terms of the Fubini-Study metric $d\Sigma_{2n}^2$ on $CP^n$ as
\be
d\Omega_{2n+1}^2 = d\Sigma_{2n}^2 + (dz+\bar{\cal A})^2\ .\label{hopfscpn}
\ee

     In fact we may give general results for any metric of the form
\be
ds^2 = c^2\, (dz + \bar{\cal A})^2 + d\bar s^2\label{bundle}
\ee
on a $U(1)$ bundle over a base manifold with metric $d\bar s^2$, where $c$
is a constant.  Choosing the vielbein basis $e^z = c \, (dz+\bar {\cal A}),
\, e^i = \bar e^i$, one finds that the Riemann tensor for $ds^2$ has
non-vanishing vielbein components given by
\bea
R_{ijk\ell } &=& \bar R_{ijk\ell} -\ft14 c^2(\bar {\cal F}_{ik}\, \bar{\cal
F}_{j\ell} - \bar{\cal F}_{i\ell}\, \bar{\cal F}_{jk}+ 2\bar{\cal F}_{ij}\,
\bar{\cal F}_{k\ell})\ ,
\nn\\
R_{zizj} &=&\ft14 c^2\, \bar{\cal F}_{ik}\, \bar{\cal F}_{jk}\ ,\qquad
R_{ijkz} = \ft12 c\, \bar\nabla_k\, \bar{\cal F}_{ij}\ .\label{curv}
\eea
In all the cases we shall consider, the components $R_{ijkz}$ will be zero,
since $\bar{\cal F}=d\bar{\cal A}$ will be proportional to
covariantly-constant tensors, such as K\"ahler forms.  The Ricci tensor for
$ds^2$ has the vielbein components
\be
R_{zz}=\ft14 c^2\, \bar{\cal F}_{ij}\, \bar{\cal F}_{ij}\ ,\qquad R_{ij}=
\bar R_{ij} 
-\ft12 c^2 \, \bar{\cal F}_{ik}\, \bar{\cal F}_{jk}\, \qquad 
R_{zi}= -\ft12 c\, 
\bar\nabla_j \bar{\cal F}_{ij}\ ,\label{kkscpn}
\ee

      Applied to our present case, where the unit $(2n+1)$-sphere 
should have a Ricci tensor satisfying $R_{ab}=2n\,\delta_{ab}$, we see that 
this is achieved by taking the field strength to be given by 
$\bar{\cal F}_{ij} 
=2 J_{ij}$, where $J_{ij}$ is the covariantly-constant K\"ahler form on 
$CP^n$.  Furthermore, the Fubini-Study Einstein metric on $CP^n$ should be 
scaled such that its Ricci tensor satisfies $\bar R_{ij}=2(n+1)\, 
\delta_{ij}$.  The volume form $\Omega_{2n+1}$ on the unit $(2n+1)$-sphere
is related to the volume form $\Sigma_{2n}$ on $CP^{n}$ by
$\Omega_{2n+1}=dz\wedge \Sigma_{2n}$.  Note also that the volume form on 
$CP^n$ is related to the K\"ahler form by
\be
\Sigma_{2n} = \ft1{n!}\, J^n\ .
\ee

\subsection{Type IIA $p$-branes}

    Let us begin by constructing the new solutions of the type IIB theory that 
are related by T-duality in the Hopf-fibred nine-dimensional background to
the string and the 5-brane of the type IIA theory.  The type IIA string
solution is given by 
\bea
ds^2 &=& H^{-3/4}\, dx^\mu\, dx_\mu + H^{1/4}\, (dr^2 + r^2\, d\Omega_7^2)
\ ,\nn\\
e^{-2\phi_1} &=& H \ , \qquad\qquad F_3^{(1)} = d^2 x \wedge d
H^{-1}= Q\, e^{\phi_1}\, \ast\Omega_7 \ ,\label{2astring} 
\eea
where $H$ is an harmonic function on the transverse space, of the form $H=1 
+ \ft16 Q\, r^{-6}$.   Using the expression (\ref{hopf7}), we may reduce
this  solution to nine dimensions in the same manner as we previously
reduced the  membrane of eleven-dimensional supergravity to $D=10$.  Thus we
obtain the  nine-dimensional solution
\bea
ds_9^2 &=& r^{2/7}\, \Big[ H^{-5/7}\, dx^\mu\, dx_\mu + H^{2/7}\, (dr^2 + 
r^2 d\Sigma_6^2) \Big]\ ,\nn\\
e^{-2\phi_1} &=& H\ , \qquad\qquad e^{2\sqrt7\phi_2} = H\, r^8\ , \nn\\
F_3^{(1)} &=& d^2x\wedge d H^{-1}=Q\, e^{\phi_1 -\ft{1}{\sqrt7}\phi_2}
\, \ast\Sigma_6\ ,\qquad\qquad {\cal F}_2^{(2)}  = 2 J\ ,\label{2a9string}
\eea
where $d\Sigma_6^2$ is the metric on $CP^3$, with K\"ahler form $J$ and
volume form $\Sigma_6$, and
$\phi_2$ is the Kaluza-Klein scalar arising in the reduction $ds_{10}^2 =
e^{-2\a\phi_2}\, ds_9^2 + e^{14\a\phi_2}\, (dz_2 +{\cal A}_1^{(2)})^2$, with
$\a=1/(4\sqrt7)$.  Note that the exponential factor appearing in the
expression for $F_3^{(1)}$ in terms of $\ast\Sigma_6$ is precisely
the inverse of the exponential dilaton prefactor of the kinetic term for
$F_3^{(1)}$ in the nine-dimensional Lagrangian. This is an example of the
general rule that a field strength with kinetic term $e^{a\phi}\, F^2$ has
the form $F=Q\, e^{-a\phi}\, \ast\Sigma$ when it carries an electric
charge. 

     Using the T-duality transformation to the type IIB
variables given in the appendix, we find that the dilatonic scalars become
$e^{-\phi}= H^{1/2}\, r$ and $e^{2\sqrt 7\varphi} = H\, r^6$, implying that
upon oxidising to $D=10$ according to $ds_{10}^2 = e^{-2\a\varphi}\, ds_9^2 +
e^{14\a\varphi} (dz_2 + {\cal A}_1)^2$, we obtain the type IIB solution 
\bea
ds_{10}^2 &=& r^{1/2}\, \Big[ H^{-3/4}\, dx^\mu\, dx_\mu + H^{1/4}\, (dr^2 + 
r^2\, d\Sigma_6^2 + r^{-2}\, dz_2^2 )\Big] \ ,\nn\\
e^{-2\phi} &=& H\, r^2\ , \qquad\qquad F_3^{(\rm NS)} = Q\, e^{\phi}\,
\ast(dz_2\wedge\Sigma_6) + 2\, J\wedge dz_2
\, \label{2a10newstring}
\eea 
Since this is related by T-duality to the string solution of the type IIA 
theory, it follows that this is a solution of the type IIB theory.  
However, as a type IIB {\it supergravity} solution, it will not 
preserve the same number of supersymmetries as it did as a solution of 
type IIA supergravity.  In fact, it will preserve either 12 or 0 of the
32 components of supersymmetry, rather than the 16 when it is a type 
IIA solution. This discrepancy  is because some of the Killing spinors
are described by stringy winding modes in the 
type IIB picture.  (The counting is the same as in the Hopf reduction 
of the M2-brane discussed in section 3.1, since in both cases the 
transverse space is foliated by 7-spheres.)   The moral to be drawn 
from this is that supergravity is not always a 
reliable guide to the number of supersymmetries preserved in string theory.

     The solution (\ref{2a10newstring}) can be viewed as a string solution 
of the type IIB theory, since it has a 2-dimensional Poincar\'e symmetry on 
the worldsheet.  However, it is a string of a rather unusual kind, with an 
unconventional geometry in the transverse space, and an unconventional 
configuration for the 3-form field strength.  However, since by construction 
it is related by T-duality to the standard string solution of the type IIA 
theory, it gives an equivalent description of these degrees of freedom.   It 
can be contrasted with two other procedures for using T-duality to relate
the type IIA string to solutions of the type IIB theory. One of these 
involves wrapping the type IIA string around a circle, giving a particle in 
$D=9$ which, after transforming to type IIB variables can be oxidised to a 
pp-wave solution of the type IIB theory in $D=10$:
\bea
ds^2 &=& -H^{-1}\, dt^2 + dr^2 +r^2\, d\Omega_7^2 + H\, (dz - H^{-1} dt)^2
\ ,\nn\\
\phi &=& 0\ .\label{2a10ppwave}
\eea
The other procedure involves first constructing a line of type IIA strings 
$D=10$, by taking $H$ to be independent of one of the Cartesian coordinates 
of the transverse space, and then compactifying along this direction.  After 
transforming this $D=9$ solution into type IIB variables, its oxidation back 
to $D=10$ will give a standard solution describing a line of NS-NS strings 
in the type IIB theory.

   We now turn to the type IIA 5-brane solution, which is given by
\bea
ds^2 &=& H^{-1/4}\, dx^\mu\, dx_\mu + H^{3/4}\, (dr^2 + r^2\, d\Omega_3^2) 
\,\nn\\
e^{2\phi_1} &=& H\ ,\qquad\qquad F_3^{(1)} = Q\, \Omega_3 \ ,\label{2a5brane}
\eea
where $\Omega_3$ is the volume form on the unit 3-sphere, and $Q$ is the 
magnetic charge carried by the 3-form field strength.
Using (\ref{hopfscpn}), we may write the 3-sphere metric as the $U(1)$ 
fibration over $CP^1\sim S^2$.  Dimensionally reducing on the $U(1)$ fibres 
gives the nine-dimensional solution
\bea
ds_9^2 &=& r^{2/7}\, \Big[ H^{-1/7}\, dx^\mu\, dx_\mu + H^{6/7}\, (dr^2 + 
r^2\, d\Sigma_2^2)\Big]\ ,\nn\\
e^{2\phi_1} &=&H\ ,\qquad\qquad e^{2\sqrt7 \phi_2} = H^3\, r^8 \ ,\nn\\
F_2^{(12)} &=& Q\, \Sigma_2\ ,\qquad\qquad {\cal F}_2^{(2)} = 2 \Sigma_2
\ ,\label{2ad95brane}
\eea
where $\Sigma_2$ is the volume form of the $CP^1$ metric, which is nothing 
but the unit 2-sphere.  Note that in this particular case the K\"ahler form 
$J$ on $CP^1$ is identical to the volume form $\Sigma_2$.
Converting to type IIB variables, and oxidising to $D=10$, we obtain the 
type IIB solution
\bea
ds^2 &=& r^{1/2}\, \Big[ dx^\mu\, dx_\mu + H\ ,(dr^2 + r^2\, d\Sigma_2^2) + 
H^{-1}\, r^{-2}\, (dz_2+{\cal A}_1)^2\Big]\ ,\nn\\
e^{-\phi} &=& r\ ,\qquad\qquad
{\cal F}_2 = Q\, \Sigma_2 \ ,\qquad\qquad F_3^{(\rm NS)} =2(dz_2+{\cal 
A}_1)\wedge \Sigma_2 \ .\label{sss}
\eea
This configuration, being related by T-duality in $D=9$ to 
the standard 5-brane of the type IIA theory, is a solution of type IIB 
supergravity.  As in the previous case of the string solution, the 
full set of 16 out of the 32 components of supersymmetry will be
preserved only once the winding modes of the type IIB string theory
are included.

     Again, as in the case of the string solution that we discussed 
previously, the type IIB solution (\ref{sss}) can be contrasted with two 
other type IIB solutions that can be obtained from the type IIA 5-brane by 
T-duality.  One of these involves a diagonal dimensional reduction of the 
type IIA 5-brane, giving a 4-brane in $D=9$.  After transforming to type IIB 
variables, this can be oxidised to $D=10$ where it describes the standard 
NS-NS 5-brane of the type IIB theory.  The other type IIB solution is 
obtained by vertical dimensional reduction of the type IIA 5-brane to a 
5-brane in $D=9$, given by
\bea
ds_9^2 &=& H^{-1/7}\, dx^\mu\, dx_\mu + H^{6/7}\, (dr^2 + r^2\, d\Omega_2^2) 
\ ,\nn\\
e^{2\phi_1} &=& H\ ,\qquad\qquad e^{2\sqrt7 \phi_2}= H^3\ ,\label{ssss}\\
F_2^{(12)} &=& Q\, \Sigma_2\ .
\eea
After transforming to type IIB variables, this oxidises to give the
NUT solution
\bea
ds^2 &=& dx^\mu\, dx_\mu + H\, (dr^2 + r^2\, d\Omega_2^2) + H^{-1}\, (dz_2 
+{\cal A}_1)^2 \ ,\nn\\
\phi &=& 0\ ,\qquad\qquad {\cal F}_2 = Q\, \Sigma_2\ \label{sssss}
\eea
of the type IIB theory in ten dimensions.

\subsection{Type IIB $p$-branes}

     The $p$-brane solutions in the type IIB string are given by
\bea
ds^2 &=& H^{-\td d/8}\, dx^\mu\, dx_\mu + H^{d/8}\, (dr^2 + r^2\, d\Omega_{\td 
d+1}^2)\ ,\nn\\
e^{\pm\phi} &=& H^{\td d/4 -1}\ ,\label{2bsol}
\eea
where the plus sign in the expression for the dilaton corresponds to 
D-branes, and the minus sign to NS-NS branes.  The $p$-brane has 
world-volume dimension $d=p+1$, and $\td d=8-d$.  The
rank $n$ of the field strength $F_n$ that supports the solution is given by
$n={\rm min}(d+1,\td d+1)$.  When $d<\td d$ the field strength carries an 
electric charge, whilst when $d>\td d$ it carries a magnetic charge:
\bea
{n=d+1}: && F_n= d^dx \wedge dH^{-1} = Q\, H^{d\td d/8-1}\, \ast\Omega_{\td 
d+1}\ ,\nn\\
{n=\td d+1}: && F_n = Q\, \Omega_{\td d+1}\ .
\label{2bfs}
\eea
In the case $d=\td d=4$, the 3-brane is supported by the self-dual 5-form
\be
F_5 = Q(\Omega_5+\ast\Omega_5) \ .
\ee
Since $\td d$ is even for all the $p$-branes, the $(\td d+1)$-sphere is 
always odd dimensional, and hence can be written in the $U(1)$-fibred form
(\ref{hopfscpn}). 

     First, let us consider the D$p$-brane solutions of the type IIB theory, 
which exist for $p=\{-1,1,3,5,7\}$.  After dimensionally reducing on the 
$U(1)$ fibre coordinate, we obtain the nine-dimensional solutions
\bea
ds_9^2 &=& r^{2/7}\, \Big[ H^{-(\td d-1)/7}\, dx^\mu\, dx_\mu + H^{d/7}\, 
(dr^2 + r^2\, d\Sigma_{\td d}^2 )\Big]\ ,\nn\\
e^{\phi}&=& H^{\td d/4-1}\ ,\qquad\qquad e^{\sqrt7 \varphi} = H^{d/4}\, 
r^4 \ .\label{2b9sol}
\eea
If the field strength $F_n$ supporting the type IIB $p$-brane carries an
electric charge, then it remains unchanged in nine dimensions, whereas if it 
carries a magnetic charge, it is reduced to $F_{n-1}$ given by 
$F_n\rightarrow F_{n-1}\wedge (dz_2 +{\cal A}_1)$, where ${\cal F}_2 = 
d{\cal A}_1 = 2J$ and $J$ is the K\"ahler form on $CP^{\td d/2}$.  After 
transforming to type IIA variables, we may then oxidise the solutions to the 
ten-dimensional type IIA theory, where we find
\bea
ds^2 &=& r^{1/2}\, \Big[ H^{-(\td d-1)/8}\, (dx^\mu\, dx_\mu + r^{-2}\, 
dz_2^2 ) + H^{(d+1)/8}\, (dr^2 + r^2\, d\Sigma_{\td d}^2) \Big] \ ,\nn\\
e^{-\phi_1} &=& H^{(d-3)/4}\, r\ .\label{2box}
\eea
The type IIA solutions are all supported by two field strengths, one of 
which is universal, namely the N-NS 3-form 
\be
F_3^{(1)} = 2Q\,  dz_2 \wedge J\ .
\ee
The other non-vanishing field strength is the R-R field strength of the type 
IIA theory whose rank $n=0$, 2 or 4 is given by $n={\rm min}(d+2,\td d)$.  
Note that the ``0-form field strength'' $F_0$ is the cosmological term of the 
massive IIA theory.  The various cases are summarised in the table below, 
where the $\hast$ symbol denotes the appropriately scaled Hodge dual 
incorporating the necessary dilaton-dependent factor $\hast = e^{-a\phi}\, 
\ast$, where the field strength has a dilaton prefactor $e^{a\phi}$ in its 
kinetic term.

\bigskip\bigskip

\centerline{
\begin{tabular}{|c|c|c|c|}\hline
 $p$ & IIB  & $D=9$ & IIA \\ \hline
$-1$ & $ d\chi = Q\, \hast\Omega_9$ & $d\chi = Q\, \hast\Sigma_8$
& ${\cal F}_2^{(1)} = Q\, \hast\Sigma_8$ \\ \hline
1 & $F_3^{\rm{R}} = Q\,\hast\Omega_7$ 
& $F_3^{\rm{R}} = Q\,\hast\Sigma_6$ & $F_4= Q\, \hast\Sigma_6$\\ \hline
3 & $F_5=Q\, (\Omega_5 + \ast\Omega_5)$ & $F_4=Q\,  \Sigma_4$ 
& $F_4 =Q\, \Sigma_4$ \\ \hline
5 & $F_3^{\rm{R}} = Q\, \Omega_3$ & $F_2^{\rm{R}} = Q\, \Sigma_2$
& ${\cal F}_2^{(1)}=Q\, \Sigma_2$ \\ \hline
7 & $d\chi=Q\, \Omega_1$ & $F_0= Q$ & $F_0=Q$\\ \hline
\end{tabular}}
\bigskip

\centerline{Table 1: Type IIB D$p$-branes and their T duals}
\bigskip\bigskip

    The configurations discussed here will be solutions of type 
IIA supergravity, but the full set of 16 out of 32 components of 
supersymmetry 
will only be found once the type IIA string winding modes are included.
For example, in the case of the type IIB 3-brane, the corresponding solution 
that we obtain by Hopf dualising on the 
fibres of the foliating 5-spheres in the transverse space will, as a 
solution of type IIA {\it supergravity}, preserve no supersymmetry at 
all.  This example emphasises the moral that just because a 
solution is  non-BPS in supergravity, it does not necessarily follow that it is 
non-BPS in string theory. (The apparent disappearance of Killing spinors under 
T-duality has also been discussed in \cite{aal,Sfetsos1}.)  

    The solutions of the type IIA theory obtained above can be 
further oxidised to $D=11$.  From (\ref{2box}), we obtain the
eleven-dimensional metrics
\bea
ds_{11}^2 &=& r^{2/3}\, \Big[ H^{d/6-1}\, 
    (dx^\mu\, dx_\mu + r^{-2}\, dz_2^2) + 
        r^{-2} H^{1-d/3}\, (dz_1 +{\cal A}_1^{(1)})^2\nn\\
&& \qquad + H^{d/6}\, (dr^2 + r^2\, d\Sigma_{\td d}^2)\Big ]
\ .\label{2box11}
\eea
The field strength $F_4$ in $D=11$ will have a universal term of the
form $2Q\, (dz_1 +{\cal A}_1^{(1)})\wedge dz_2\wedge J$, together with an
extra term of the form given in Table 1, in the cases $p=1$ or $p=3$.
The Kaluza-Klein vector ${\cal A}_1^{(1)}$ will be zero when $p=1$ or
3, while $d{\cal A}_1^{(1)}$ will give the field strengths ${\cal
F}_2^{(1)}$ listed in Table 1 when $p=-1$ or $p=5$.  The $p=7$ case
is a solution of the massive IIA theory, and presumably cannot be 
oxidised to $D=11$.

    Note that in the case of $p=3$, we find from (\ref{2box11}) that
the Hopf dualisation of the type IIB self-dual 3-brane gives the
eleven-dimensional solution
\bea
ds_{11}^2 &=& r^{2/3}\, \Big[ H^{-1/3}\, 
    (dx^\mu\, dx_\mu + r^{-2}\, (dz_1^2+ dz_2^2))  
       + H^{2/3}\, (dr^2 + r^2\, d\Sigma_4^2)\Big ]\ ,\nn\\
F_4&=& 2Q\, dz_1\wedge dz_2 \wedge J + Q\, \Sigma_4\ .
\label{2b3box11}
\eea

    In addition to the D-branes discussed above, the type IIB theory
also has NS-NS string and 5-brane solutions.  The T-duality
transformation of these into solutions of the type IIA theory
proceeds, {\it mutatis mutandis}, identically to the discussion of the
transformation to type IIB of the type IIA string and 5-brane in the
previous subsection.  This follows from the fact that the NS-NS sector
of the two type II theories are identical, and invariant under the
T-duality.

\section{AdS$_5\times S^5$ compactification of type IIB}

     As we discussed in section 3, certain extremal $p$-brane
solutions, namely those where the dilaton is finite on the horizon,
have a spacetime structure that approaches AdS$\times S^n$ as the
horizon of the $p$-brane is approached.  In this section, we shall
consider a particular such example, namely the AdS$_5\times S^5$
solution of the type IIB theory, which may be viewed as the
near-horizon limit of the self-dual 3-brane.  For the present
purposes, however, we shall find it more convenient not to obtain the
solution by this limiting process, but rather, to work directly with
the AdS$_5\times S^5$ solution.

      This solution involves just the metric tensor and the self-dual
5-form field strength $H_\5$ of the type IIB theory, whose relevant
equations of motion can be written simply as
\bea
R_{MN} &=& \ft1{96}\, H_{MPQRS}\, H_N{}^{PQRS}\ ,\nn\\
H_\5 &=& {* H_\5}\ ,\label{gheq}
\eea
where, in the absence of the other fields of the theory, we have
simply $H_\5=dB_\4$.  We may find a solution on AdS$_5\times S^5$ of
the form
\bea
ds^2 &=& ds^2(AdS_5) + ds^2(S^5)\ ,\nn\\
H_\5 &=& 4m \Omega_{AdS_5} + 4m \Omega_{S^5}\ ,\label{2bconfig}
\eea
where $\Omega_{AdS_5}$ and $\Omega_{S^5}$ are the volume forms on
AdS$_5$ and $S^5$ respectively, $m$ is a constant, and the metrics on
AdS$_5$ and $S^5$ satisfy
\be
R_{\mu\nu} = -4 m^2\, g_{\mu\nu}\ ,\qquad 
R_{mn} = 4 m^2\, g_{mn}
\ee
respectively.  Since the unit 5-sphere has metric $d\Omega^2_5$
with Ricci tensor $\bar R_{mn} = 4\, \bar g_{mn}$, it follows that we 
can write 
\be
ds^2(S^5) = \ft1{m^2}\, d\Omega^2_5\ .
\ee
From (\ref{hopfscpn}), it follows that we can write this as 
\be
ds^2(S^5) = \ft1{m^2}\, d\Sigma_4^2 + \ft1{m^2}\, 
(dz+\bar{\cal A})^2\ ,
\ee
where $d\Sigma_4^2$ is the metric on the ``unit'' $CP^2$, and $d
\bar{\cal A}= 2J$, where $J$ is the K\"ahler form on $CP^2$.

     We may now perform a dimensional reduction of this solution to
$D=9$, by compactifying on the circle of the $U(1)$ fibres,
parameterised by $z$.  Comparing with the general Kaluza-Klein
prescription, for which
\bea
ds_{10}^2 &=& ds_9^2 + (dz_2 +{\cal A})^2\ ,\nn\\
H_\5 &=& H_\5 + H_\4\wedge (dz_2 + {\cal A})\ ,
\eea
we see, from the fact that the $S^5$ and $CP^2$ volume forms are
related by $\Omega_5 = (dz+\bar{\cal A})\wedge \Sigma_4$, that the
solution will take the 9-dimensional form
\bea
ds_9^2 &=& ds^2(AdS_5) + \ft1{m^2}\, d\Sigma_4^2\ ,\nn\\
F_\4 &=& \ft4{m^3}\, \Sigma_4\ ,\qquad {\cal F}_\2 = \ft2{m}\, J\ .
\label{d92bsol}
\eea
(Note that in the dimensional reduction of the 5-form of the type IIB
theory, its self-duality translates into the statement that the fields
$H_\5$ and $H_\4$ in $D=9$ must satisfy $H_\4 ={* H_\5} =F_\4$.)

     We now perform the T-duality transformation to the fields of the
$D=9$ reduction of the type IIA theory.  The relation between the IIB
and the IIA fields is given in the appendix.  Thus in the IIA
notation, we have the nine-dimensional configuration
\bea
ds_9^2 &=& ds^2(AdS_5) + \ft1{m^2}\, d\Sigma_4^2\ ,\nn\\
F_\4 &=& \ft4{m^3}\, \Sigma_4\ ,\qquad F_2^{(12)} = \ft2{m}\, J\ .
\label{2asol}
\eea
The crucial point is that the 2-form field strength $F_2^{(12)}$ of
the IIA variables is no longer a Kaluza-Klein field coming from the
metric; rather, it comes from the dimensional reduction of the 3-form
field strength in $D=10$.  Indeed, if we trace the solution
(\ref{2asol}) back to $D=10$, we have the type IIA configuration
\bea
ds_{10}^2 &=& ds^2(AdS_5) + \ft1{m^2}\, d\Sigma_4^2 + dz_2^2\ ,\nn\\
F_\4 &=& \ft4{m^3}\, \Sigma_4\ ,\qquad F_3^{(1)} = \ft2{m}\, J\wedge dz_2\ .
\label{2ad10}
\eea
The solution has the topology AdS$_5\times CP^2\times S^1$.  This
should be contrasted with the topology AdS$_5\times S^5$ for the
original $D=10$ solution in the type IIB framework.  Thus the
T-duality transformation in $D=9$ has ``unravelled'' the twisting of
the $U(1)$ fibre bundle over $CP^2$, leaving us with a direct product
$CP^2\times S^1$ compactifying manifold in the type IIA description.

   A further oxidation to $D=11$ can now be performed.  Upon doing so,
we obtain the configuration
\bea
ds_{11}^2&=& ds^2(AdS_5) + \ft1{m^2}\, d\Sigma_4^2 + dz_1^2 +dz_2^2\ ,\nn\\
F_\4 &=& \ft4{m^3}\, \Sigma_4 - \ft2{m}\, J\wedge dz_1\wedge dz_2\ ,
\label{2ad11}
\eea
which is just the near-horizon limit of (\ref{2b3box11}).
The topology of this solution is AdS$_5\times CP^2\times T^2$.

     At first sight, the T-duality transformation that we have
performed has a somewhat surprising implication.  We began with a
solution on AdS$_5\times S^5$, which admits a spin structure, and
mapped it {\it via} T-duality to a solution on AdS$_5\times
CP^2\times S^1$, which does not admit a spin structure (because $CP^2$
does not admit a spin structure).  In particular, this means that the
spectrum of Kaluza-Klein excitations in the $CP^2\times S^1$
compactification of type IIA supergravity contains no fermions at all!

    To understand this, we should first look at the situation in the
type IIB language, after having performed the reduction on the circle
of the $U(1)$ fibres, but before we make the T-duality transformation
to the type IIA fields.  Here too, we have a compactification
involving $CP^2$, namely the AdS$_5\times CP^2$ solution of $D=9$
supergravity.  However, at this stage we have done nothing but
re-write the AdS$_5\times S^5$ solution in terms of reduced $D=9$
fields.  The crucial point is that in this description, given in
(\ref{d92bsol}), the Kaluza-Klein potential ${\cal A}$ has a
topologically non-trivial form, with its field strength being
proportional to the K\"ahler form of $CP^2$.  All the fermions in the
Kaluza-Klein expansion of the $D=10$ type IIB fermions will be charged
with respect to this Kaluza-Klein potential.  As discussed in
\cite{hawkingpope}, $CP^2$ does admit a spin$^c$ structure, or
generalised spin structure, in which the spinors are charged under the
gauge potential whose field strength is the K\"ahler form $J$.  In
fact the non-existence of a standard spin structure is caused by the
fact that spinors transported around a family of closed curves
spanning the non-trivial 2-cycle in $CP^2$ differ in phase by a
factor of $-1$, implying an inconsistency.  This inconsistency is
removed by considering instead {\it charged} spinors, minimally
coupled to the gauge potential for $J$, whose charges $q$ are chosen
to be precisely intermediate between the values that would normally be
required by the Dirac quantisation condition in the presence of the
magnetic charge $\int J$; in other words of the form $q=n+\ft12$.
Normally, this would give a minus sign inconsistency in the fermion
phases, but here it precisely cancels the previously-discussed
minus-sign inconsistency, allowing the existence of the charged
spinors.  Under the dimensional reduction on the $U(1)$ fibres of
$S^5$, all the fermions automatically acquire proper charges that are
consistent with the generalised spin structure.  (This is discussed in
some detail in \cite{pope}.)  Thus in the type IIB description, there
is a complete consistency between the $D=10$ and $D=9$ pictures.

    Now, let us perform the $T$-duality transformation to the
nine-dimensional type IIA field variables.  In particular, this means
that the non-trivial 1-form potential will no longer be a Kaluza-Klein
vector potential, but instead it is the winding-mode potential
$A_1^{(12)}$ coming from the dimensional reduction of the NS-NS 2-form
$A_2^{(1)}$ in $D=10$ type IIA.  The Kaluza-Klein modes in $D=9$ type
IIA do not carry charges with respect to this potential.
Consequently, Kaluza-Klein fermions in $D=9$ could not evade the
sign-inconsistency problem that precludes the existence of a spin
structure in $CP^2$.  In other words, there can be no fermions at all
in the spectrum of Kaluza-Klein excitations of the $D=10$ type IIA
theory compactified on $CP^2\times S^1$.  In the string theory
there are, however, also
winding modes to be considered.  These modes {\it are} charged with
respect to the winding mode potential $A_1^{(12)}$, and it is fermions
in this sector of the complete $D=9$ spectrum that will carry the
necessary charges that allow them to exist consistently on $CP^2$.

     We are now in a position to consider in more detail the relation
between the spectrum of states in the AdS$_5\times S^5$
compactification of the type IIB theory, and the AdS$_5\times
CP^2\times S^1$ compactification of the type IIA theory.  The easiest
way to describe this is by looking first at the states in type IIB,
which carry representations of the $SO(6)=SU(4)$ isometry group of
$S^5$, and decompose them with respect to the $SU(3)\times U(1)$
subgroup which is the isometry group of $CP^2$ times the $U(1)$ gauge
symmetry of the Kaluza-Klein gauge potential.  In particular, the 8
gravitini and the 15 gauge bosons of $SO(6)$ decompose as
\bea
15 &\longrightarrow& 8_0 + 1_0 + 3_{-2} + \bar 3_{2}\ ,\nn\\
4+ \bar 4 &\longrightarrow& 3_{-1/2} + 1_{3/2} + \bar 3_{1/2} + 1_{-3/2}\ .
\label{decomp}
\eea 
The subscripts on the $SU(3)$ representations denote their charges
with respect to $U(1)$.  These are the Kaluza-Klein charges $q$,
associated with the dependence $e^{iqz}$ on the compactifying
coordinate $z$.  In particular, we see that the gravitini have
non-zero charges, and so all are truncated out in a dimensional
reduction to the $z$-independent sector.  Furthermore, as we mentioned
previously, they have half-integer charges, precisely as is needed for
consistency in the $CP^2$ manifold.

     Turning to the gauge bosons, we see that the $8+1$ gauge bosons
of the $SU(3)$ isometry of $CP^2$, together with the $U(1)$
Kaluza-Klein gauge potential, survive in a truncation to the zero-mode
sector.  The rest of the $SO(6)$ gauge symmetry of the 5-sphere is
recovered only if the charged Kaluza-Klein modes are
retained.\footnote{Analogous considerations of the Hopf fibration of
$S^7$ in the context of AdS$_4\times S^7$ compactifications of $D=11$
supergravity and M-theory were explored in \cite{nilpop,DLP}.}

     After dualising to the $D=9$ type IIA picture, the $U(1)$ charges
in (\ref{decomp}) will be carried instead by the gauge potential
$A_1^{(12)}$ coming from the dimensional reduction of the 2-form
potential $A_2^{(1)}$ of the type IIA theory.  Thus the charged
representations in (\ref{decomp}) will not be seen at all in the
Kaluza-Klein spectrum of the compactification of type IIA supergravity
on AdS$_5\times CP^2\times S^1$.  It is only by including the winding
modes of the type IIA string that the charged representations will be
recovered.  (Subtleties involving the interpretation of T-duality as
the interchange of Kaluza-Klein modes and winding modes in cases such
as we are considering, where there is a $U(1)$ isometry but no
non-contractible loop, are discussed in \cite{aal}.  We shall continue
to use the term ``winding mode,'' even though it is perhaps not wholely
appropriate in this context.)

    Note that although the type IIA Kaluza-Klein spectrum has states of
maximum spin 2, these states are not BPS, and indeed will belong to long
supermultiplets of the type IIA string when the winding modes are
included.

    Finally, we address the puzzle that on the type IIA side the
gauge symmetry is $SU(3)\times U(1)\times U(1)\times U(1)$, but only
an $SU(3)\times U(1)$ sits inside the $SO(6)$ of the type IIB side.
(Two of the $U(1)$ factors are the obvious Kaluza-Klein $U(1)$'s from
the $T^2$ compactification of M-theory, while the third is associated
with the gauge potential ${\cal A}_1^{(12)}$; see Table 4 in the Appendix.)
The R-R and NS-NS vectors ${\cal A}_1^{(i)}$ which form an $SL(2,Z)$ 
doublet of $D=9$ type IIA supergravity survive, after compactification
on $CP^2$, but in the type IIB interpretation belong to ``massive''
multiplets of $N=8$, $D=5$ supersymmetry.  (That is to say, they are
not in the massless supergravity multiplet \cite{KRV}.)

\section{Non-maximally-supersymmetric compactifications}

    The AdS$_5\times S^5$ solution of the previous section can be
generalised to any other configuration of the form AdS$_5\times M_5$,
where $M_5$ is any five-dimensional compact Einstein space with
positive Ricci tensor.  A particularly interesting class of such
solutions is provided by taking $M_5$ to be a $U(1)$ bundle over
$S^2\times S^2$.  We may denote these spaces by $Q(n_1,n_2)$, where
the integers $n_1$ and $n_2$ are the winding numbers of the fibres
over the two $S^2$ factors in the base manifold.  Natural metrics on
these spaces are given by \cite{pagepope}
\be
ds^2 = c^2\, (dz + \bar{\cal A})^2 +
\fft1{\Lambda_1} \, (d\theta_1^2 + \sin^2\theta_1\, d\phi_1^2) 
+\fft1{\Lambda_2} \, (d\theta_2^2 + \sin^2\theta_2\, d\phi_2^2)
\ ,\label{s2s2}
\ee
where $c$ is a constant, $\Lambda_1$ and $\Lambda_2$ are the
``cosmological constants'' of the two 2-spheres in the base manifold,
and $z$ has period $2\pi$. The potential $\bar{\cal A}$ can be taken
to be
\be
\bar{\cal A}= -n_1\, \cos\theta_1\, d\phi_1 - n_2\, \cos\theta_2\,
d\phi_2 \ ,
\ee
giving a field strength with the vielbein components
\be
\bar{\cal F}_{i_1 j_1} = n_1\, \Lambda_1\, \epsilon_{i_1 j_1}\ ,\qquad
\bar{\cal F}_{i_2 j_2} = n_2\, \Lambda_2\, \epsilon_{i_2 j_2}\ ,\label{epep}
\ee
where $\epsilon_{i_1 j_1}$ and $\epsilon_{i_2 j_2}$ are the Levi-Civita 
tensors on the two 2-spheres.
Substituting into (\ref{kkscpn}), we see find that the Ricci tensor for
$ds^2$ has the vielbein components
\bea
&&R_{i_1j_1} = (\Lambda_1 -\ft12n_1^2\,  c^2\, \Lambda_1^2)\, \delta_{i_1j_1}
\ ,\qquad
R_{i_2j_2} = (\Lambda_2 -\ft12n_2^2\,  c^2\, \Lambda_2^2)\, \delta_{i_2j_2}
\ ,\nn\\
&&R_{zz} = \ft12 c^2\, (n_1^2\, \Lambda_1^2 + n_2^2\, \Lambda_2^2)\ ,
\label{s2s2ricci}
\eea
It is easy to see that for each choice of integers $n_1$ and
$n_2$, there are uniquely determined quantities $x_1$ and $x_2$ such that  
the metric $ds^2$ is Einstein, with $\Lambda_1=x_1\, c^{-2}$ and
$\Lambda_2= x_2\, c^{-2}$ \cite{pagepope}.  In fact $x_1$ and $x_2$
are the real roots of the cubic polynomials
\bea
9 n_1^4\, n_2^2\, x_1^3 - 24 n_1^2 \, n_2^2\, x_1^2 + 8(n_1^2+2 n_2^2)\, x_1
-8&=&0\ ,\nn\\
9 n_2^4\, n_1^2\, x_2^3 - 24 n_2^2 \, n_1^2\, x_2^2 + 8(n_2^2+2 n_1^2)\, x_2
-8&=&0\ .\label{cubic}
\eea  
Thus we have an Einstein metric for each $Q(n_1,n_2)$ space.  We may
take $n_1$ and $n_2$ to be relatively prime, since if they had a
common divisor, it would simply imply that the period for $z$ could
have been taken to be $2\pi$ times gcd$(n_1,n_2)$, rather than simply
$2\pi$.  Thus if we always take $z$ actually to have the period
$2\pi$, the $Q(n_1,n_2)$ spaces where $n_1$ and $n_2$ have a common
divisor are ``lens spaces,'' where the fibres have been identified
under the freely-acting group $Z_p$, where $p=\hbox{gcd}(n_1,n_2)$.
In fact, we have in general that $Q(n_1,n_2) = Q(n_1/p,n_2/p)/Z_p$.

     The situation of greatest interest to us is when the Einstein
space $M_5$ admits Killing spinors, implying that the AdS$_5\times
M_5$ solution will preserve some supersymmetries. There will be an
unbroken supersymmetry for each solution $\eta$ of the Killing-spinor
equation on $M_5$.  In this AdS context the Killing-spinor equation is
\cite{DNP}
\be
{\cal D}_a\, \eta \equiv D_a \,\eta -\ft{i}2 m\, \Gamma_a\,\eta\
=0\ ,\label{ksp}
\ee
where we assume that the Einstein metric on $M_5$ is such that $R_{ab}
= 4m^2 \, \delta_{ab}$ (in vielbein components).  The integrability
condition for the existence of solutions to (\ref{ksp}), obtained by
taking the commutator of the ${\cal D}_a$ derivatives, is
\be
{[} {\cal D}_a, {\cal D}_b {]}\eta  = \ft14 R_{abcd}\, \Gamma^{cd}\, \eta 
-\ft12 m^2\, \Gamma_{ab}\, \eta =0\ .\label{intcon}
\ee

    We find that the Einstein metrics on $Q(n_1,n_2)$ admit Killing
spinors only if $n_1=n_2=1$.  In this case, solving the Einstein
conditions following from (\ref{s2s2ricci}), we obtain an Einstein
metric on $Q(1,1)$ for which $R_{ab} =4m^2\, \delta_{ab}$, provided
that the parameters in (\ref{s2s2}) are given by
\be
\Lambda_1=\Lambda_2 = 6m^2\ ,\qquad c=\fft1{3 m} \ .
\ee
It is straightforward now to substitute the resulting expressions for
the Riemann tensor, given by (\ref{curv}), into the integrability
condition (\ref{intcon}).  We find that Killing spinors exist provided
that they satisfy the condition
\be
\Gamma_{1234}\, \eta =\eta\ ,
\ee
where 1 and 2 are the vielbein indices for the first 2-sphere, and 3
and 4 are the indices for the second 2-sphere.  Thus there are half
the number of Killing spinors on $Q(1,1)$ as there are on $S^5$, and
so we have a solution AdS$_5\times Q(1,1)$ with $N=4$ spacetime
supersymmetry.  The isometry group of $Q(1,1)$ is $SO(4)\times U(1) 
=SU(2)\times SU(2)\times U(1)$.\footnote{Curiously enough, the 
topology of the space $Q(1,1)$
is in fact $S^2\times S^3$, although the Einstein metric is not the
standard direct-product metric on $S^2\times S^3$ \cite{ziller}.  The
space $Q(1,1)$ is the coset $SO(4)/SO(2)$, which is the $S^2$ bundle
of unit tangent vectors over $S^3$.  This bundle is trivial, since
$S^3$ is parallelisable, which explains why the topology of $Q(1,1)$
is just the direct product $S^2\times S^3$.  (The spaces $Q(0,1)$ and
$Q(1,0)$ also have topology $S^2\times S^3$, but give rise to the
``standard'' direct-product Einstein metric for this topology; in
these cases there will be no Killing spinors.)}  This corresponds to
one of the gauged $N=4$, $D=5$ theories described in \cite{romans}. 
The massive Kaluza-Klein modes will in general have irrational AdS$_5$
energies in this compactification, implying that we would need to take
the covering space of AdS$_5$ where the time coordinate is non-compact.
Alternatively, if the AdS$_5$ spacetime is still required to
have its usual periodic time coordinate then the the AdS energies will
necessarily be integer or half-integer, implying that only a subset of
the Kaluza-Klein modes will survive.

    Whenever an Einstein space that is not a round sphere has Killing
spinors, its orientation reversal gives a space with no Killing
spinors \cite{DNP}.  (Except for round spheres, the equation
(\ref{ksp}) admits solutions only for one choice of sign of $m$, and
by reversing the orientation of the manifold, (\ref{ksp}) admits
solutions for the ``wrong sign,'' implying that there are no spacetime
supersymmetries.)  By this means we can also obtain a compactification on
AdS$_5\times Q(1,1)$ that has no supersymmetry. 

     We may now follow steps analogous to those described for $S^5$ in
the previous section, and reduce the AdS$_5\times Q(1,1)$ solution of
the type IIB theory to $D=9$, and perform a T-duality transformation.
Upon oxidation back to the $D=10$ type IIA theory, we have a solution
on AdS$_5\times S^2\times S^2 \times S^1$.  This can be oxidised
further to $D=11$, giving a solution on AdS$_5\times S^2\times S^2
\times T^2$.

    There are also other ways in which we may obtain AdS$_5\times M_5$
compactifications with less than maximal supersymmetry.  In
particular, we may take the standard AdS$_5\times S^5$ solution, and
simply replace $S^5$ by the cyclic lens space of order $k$, obtained
by identifying the fibre coordinate of the $U(1)$ bundle over $CP^2$
with a period which is $1/k$ times the period in the $S^5$ case.  We
may denote these lens spaces by $S^5/Z_k$.  (This mechanism for
reducing the supersymmetries was proposed in \cite{DNP3}.) It is
evident that the
mode functions on $S^5/Z_k$ will be the subset of mode functions on
$S^5$ whose $U(1)$ charges $q$ are of the form
\be
q=\ft12 k\, n\ ,\label{subset}
\ee
where $n$ is any integer.  Thus the spectrum of Kaluza-Klein states in
AdS$_5$ will now be given by this subset of states of the AdS$_5\times
S^5$ compactification.  All that is necessary in order to determine
what survives is to take the decompositions of all the $SO(6)$
representations under the $SU(3)\times U(1)$ subgroup, and retain only
those whose $U(1)$ charges satisfy (\ref{subset}).  For example, we can
see from (\ref{decomp}) that if we consider the lens space $S^5/Z_3$,
only the two $SU(3)$ singlet gravitini will survive, and so the $N=8$
supersymmetry of the $S^5$ compactification will be broken to $N=2$.
At the same time, only the gauge bosons of $SU(3)\times U(1)$ will
survive.  The $U(1)$ gauge boson is in the $N=2$ supergravity
multiplet, and the $SU(3)$ gauge bosons will be in matter 
multiplets.\footnote{This example is the same as one discussed
in \cite{Kachru,lnv}, which was obtained by describing $S^5$ as the unit
sphere in $C^3$, and then identifying the three complex
coordinates $z^i$ under $z^i\rightarrow e^{i\alpha}\,z^i$, with
$\alpha=2\pi/3$.  In fact, from the construction of the Fubini-Study
metric on $CP^2$ as the Hopf fibration of $S^5$, one can see that
making such an identification for {\it any} $\alpha=2\pi /k$ will give
rise to the cyclic lens space $S^5/Z_k$.  (See, for example,
\cite{gibbpop}.)}  Another possibility is to consider the lens space
$S^5/Z_k$ for any other values of $k$ apart from $k=1$ or $k=3$.  Now,
we see from (\ref{decomp}) that none of the gravitini (nor indeed any
fermions at all if $k$ is even) will survive, and so we obtain 
$N=0$ solutions.
These are different from the $N=0$ theory obtained in \cite{Kachru}.
Note that in all these example, since the surviving states are a
subset of the original states on $S^5$, their AdS$_5$ energies, and
hence the conformal weights of the associated operators in the
Yang-Mills theory, will all be of standard integer form.  This is
quite different from the situation for the $Q(1,1)$ compactification
that we described previously (since generic Kaluza-Klein
compactifications will give rise to fractional, and indeed irrational,
mass eigenvalues \cite{DNP}).  

\section{Further examples}

       Non-dilatonic $p$-branes are defined to be either $p$-branes
where there is no dilaton coupling, or those where the dilatons are
regular on the horizon.  Consequently, these $p$-branes have the
common feature that their metrics are regular on the horizon.  In
particular, the metrics have the form of AdS$_{p+2}\times S^{D-p-2}$.
Thus they can be viewed as interpolating between AdS$_{p+2}\times
S^{D-p-2}$ on the horizon and $D$-dimensional Minkowski spacetime
asymptotically at infinity.  The M-theory membrane and 5-brane in
$D=11$, and the self-dual 3-brane in the type IIB theory, are the
three examples, which we discussed earlier.  These $p$-branes are
supported by a single field strength which carries either a single
electric or magnetic or self-dual charge, and the solutions involve a
single harmonic function.  Non-dilatonic $p$-branes also exist in
lower dimensions, for example the dyonic strings in $D=6$ 
\cite{Rahmfeld},
three-charge black holes and strings in $D=5$ \cite{Tseytlin1}, and 
four-charge black
holes in $D=4$ \cite{Cvetic1,Cvetic2}.  Upon oxidising these solutions
to $D=11$ or $D=10$, they become intersections of $p$-branes, waves and NUTs.

        We shall study the horizons of these non-dilatonic $p$-branes
as solutions in their own right.  Since the dilatonic scalars decouple
({\it i.e.}\ they are constants), it follows that the oxidation of
these metrics to $D=11$ must be described by AdS$_{p+2} \times
S^{D-p-2} \times T^{11-D}$, where $T^{11-D}$ is an
$(11-D)$-dimensional torus.\footnote{Alternatively, some or all of the
torus directions could be taken to be non-compact, so that $T^{10-D}$ 
would be replaced by $T^{m_1}\times E^{m_2}$, where $m_1+m_2 =10-D$.
Note also that there are solutions where $T^{10-D}$ is replaced by any
Ricci-flat $(10-D)$-dimensional space.} 
Oxidising instead to $D=10$, the metric
must be of the form AdS$_{p+2}\times S^{D-p-2} \times T^{10-D}$,
independent on whether it is oxidised to the type IIA or the type IIB
theory. (Here we consider lower-dimensional solutions that are
supported by field strengths come from the dimensional reduction of
antisymmetric tensors from $D=11$ or $D=10$.  We comment on the cases
later where the solutions are supported by field strengths that come
from the dimensional reduction of the metric.)  Thus to summarise,
M-theory, type IIA strings and type IIB strings have the following
solutions that are of AdS structure, namely
\bigskip\bigskip

\centerline{
\begin{tabular}{|c|c|c|}\hline
M-theory & type IIA & type IIB \\ \hline
AdS$_2\times S^2 \times T^7$ & AdS$_2\times S^2 \times T^6$ &
AdS$_2\times S^2 \times T^6$ \\
AdS$_3\times S^2 \times T^6$ & AdS$_3\times S^2 \times T^5$ &
AdS$_3\times S^2 \times T^5$ \\
AdS$_2\times S^3 \times T^6$ & AdS$_2\times S^3 \times T^5$ &
AdS$_2\times S^3 \times T^5$  \\
AdS$_3\times S^3 \times T^5$ & AdS$_3\times S^3 \times T^4$ &
AdS$_3\times S^3 \times T^4$  \\
AdS$_4\times S^7$            &                               & \\
 & & AdS$_5\times S^5$             \\
AdS$_7\times S^4$            &                               &
                              \\ \hline
\end{tabular}}

\bigskip

\centerline{Table 2:  AdS and sphere structures in type IIA,
IIB and M-theory}
\bigskip

      Note that the solutions that can be viewed as the horizons of
M-branes or self-dual 3-branes that are supported by either the 4-form
field strength or the self-dual 5-form, in a uniquely determined way;
the 4-form field strength in $D=11$ is given by the volume forms of
the AdS$_4$ or $S^4$ respectively and the self-dual 5-form in $D=10$
is given by the sum of the volume forms of the AdS$_5$ and $S^5$.  The
other solutions, which can be viewed as the horizons of intersecting
branes, can be supported by different field strengths.  To see this,
let us consider some examples.  In $D=6$, a dyonic string can be
supported by a field strength $F_3^{(i)}$, which carries both electric
and magnetic charges.  Its horizon is AdS$_3\times S^3$.  Oxidising
this back to $D=11$, we obtain the metric AdS$_3\times S^3\times T^5$,
which is the horizon of the intersection of a membrane and a 5-brane,
and the solution is unique.  If we oxidise the solution to $D=10$ type
IIA, different situations can arise depending on the value of the
internal index $i$.  If $i=1$, then the solution will become the
horizon of the intersection of a string and a 5-brane, supported by
the NS-NS 3-form; if $i=2, 3, 4$ or 5, then the solution will become
the horizon of the intersection of a D2-brane and a D4-brane,
supported by the R-R 4-form.  Now the situation is more complicated if
we oxidise the solution to type IIB.  If $i=1$, the result is the same
as in type IIA, namely it becomes the horizon of the NS-NS string and
5-brane; if $i=2$, then it becomes the intersection of the D-string
and D5-brane, supported by the R-R 3-form field strength; if $i=3,4$
or 5, then it becomes intersection of two D3-branes.  This example
shows that although the metrics for all these solution have the same
form, namely AdS$_3\times S^3\times T^5$, they can be supported by
quite different types of field strengths.  Another way of obtaining 
the geometries in Table 2 is by performing duality transformations on 
intersecting brane solutions \cite{hyun,Boonstra1,bergbehr,cllpst}.

        The situation is analogous, but more complicated for three-charge
and four-charge non-dilatonic $p$-branes.  The possible field
strengths that can be used to construct such a solution are given by
\cite{classp}
\bea
D=5:&& \{F_2^{(ij)},F_2^{(k\ell)},F_2^{(mn)}\}_{15}\ ,\qquad
       \{*F_3^{(i)},{\cal F}_2^{(j)}, F_2^{(ij)}\}_{30}\ ,
\label{d5sols}\\
D=4:&& \{F_2^{(ij)},F_2^{(k\ell)},F_2^{(mn)},*{\cal
             F}_2^{(p)}\}_{105+105}\ ,\qquad
\{ F_2^{(ij)}, *F_2^{(ik)}, {\cal F}_2^{(j)}, {\cal
F}_2^{(k)}\}_{210}\ ,\nn\\
&& \{F_2^{(ij)},F_2^{(k\ell)}, *F_2^{(ik)}, *F_2^{(j\ell)}\}_{210}
\ ,\label{d4sols}
\eea
where the indices $(i,j,\ldots)$ are all different, and run over all
the internal dimensions.  The field strengths with $*$ and without $*$
carry electric and magnetic charges respectively, or else magnetic and
electric charges.  The subscripts denote the multiplicities of the
solutions.  These solutions form 45- and 630-dimensional
representations of the Weyl group of $E_6$ and $E_7$ \cite{lpsweyl}.
Although different solutions have the same metric configuration, they
are supported by quite different field strengths.  In particular, it
implies that the AdS$_2$ or AdS$_3$ solutions in type IIA, type IIB or
M-theory can have very different field-strength configurations.  For
example, the black hole solution supported by the field strengths
$\{F_2^{(13)}, F_2^{(24)}, F_2^{(56)}\}$ becomes the intersection of
three membranes in $D=11$, or one string and two membranes in type
IIA, or one NS-NS string, one D-string and one D3-brane in type IIB.
The classification of such correspondences between lower-dimensional
solutions and higher-dimensional intersections can be found in
\cite{classp}.   The various field strengths appearing in
(\ref{d5sols}) and (\ref{d4sols}) divide between NS-NS and R-R as
follows:
\bea
\hbox{NS-NS}:&& F_3^{(1)}\ ,\qquad F_2^{(1\a)}\ ,
                             \qquad {\cal F}_2^{(\a)}\ ,\nn\\
\hbox{R-R}:&& F_3^{(\a)}\ ,\qquad F_2^{(\a\beta)}\ ,\qquad
              {\cal F}_2^{(1)}\ .\label{nsrr}
\eea

     There is one more example that is worth mentioning.
In $D=4$, one can construct dyonic black holes if a single 2-form
field strength carries both electric and magnetic charges.  The
solution is non-supersymmetric and describes a bound state with
negative binding energy \cite{gk}.  The horizon of this solution is
AdS$_2\times S^2$, and hence the supersymmetry is fully restored at
the horizon.  This provides further field configurations that support
AdS structures in supergravities.

          In the above discussion, we have considered lower
dimensional solutions that are supported by the field strengths that
come from the dimensional reduction of antisymmetric tensors in $D=11$
or $D=10$.  In these cases, upon oxidation, the metric is still
diagonal, and the AdS structure is manifest.  As we have seen in
(\ref{d5sols}) and (\ref{d4sols}), there can also be lower-dimensional
solutions that are supported by Kaluza-Klein 2-form field strengths,
coming from the dimensional reduction of the metric.  They can be used
to construct Riessner-Nordstr{\o}m black holes in $D=5$ and $D=4$,
which approach AdS$_2$ near the horizon.  Upon oxidation, the metric
acquires off-diagonal components, and describes a gravitational wave.
The form of the metric near the horizon is not AdS$_2\times S^2 \times
T^n$, as it would be for those examples in the first list in
(\ref{d5sols}) and the third list in (\ref{d4sols}).  In fact the
near-horizon form of the higher-dimensional metric in this case
approaches a metric which is locally AdS$_3\times$sphere$\times$torus
\cite{KSKS,DBISSS,VBFL}.

      In the above AdS and sphere solutions of the type IIA, type IIB
and M theories, there are further examples where the sphere has odd
dimension, namely $S^3$, which can be viewed as a $U(1)$ bundle over
$CP^1=S^2$.  Thus for such a type IIA or type IIB solution with $S^3$
in $D$ dimensions, we can perform a Hopf T-duality transformation on
the $U(1)$ coordinate, upon reducing to $(D-1)$ dimensions.  If the
solution is supported by R-R fields, then this transformation will
have the effect of untwisting the $S^3$ to give a solution on
$CP^1\times S^1$ in $D$ dimensions.  Note that the T-duality in
question here is part of the T-duality {\it symmetry} of the
$(D-1)$-dimensional theory.  
If these metrics are in the type IIA theory, they can then be further
mapped to M-theory with $T^2 \times CP^1$.  Thus we can summarise the
AdS and $CP^n$ structures as follows:
\bigskip\bigskip

\centerline{
\begin{tabular}{|c|c|c|}\hline
M-theory & type IIA & type IIB \\ \hline
AdS$_2\times CP^1 \times T^7$ & AdS$_2\times CP^1 \times T^6$ &
AdS$_2\times CP^1 \times T^6$ \\
AdS$_3\times CP^1 \times T^6$ & AdS$_3\times CP^1 \times T^5$ &
AdS$_3\times CP^1 \times T^5$ \\
AdS$_5\times CP^2 \times T^2$ &  &  \\ \hline
\end{tabular}}

\bigskip

\centerline{Table 3:  AdS and $CP^n$ structures in type IIA,
IIB and M-theory}
\bigskip

          So far we have looked at the AdS structures in maximal
supergravities.  Such structures also exist in non-maximal
supergravities.  For example, in the AdS solutions listed in Table 2
and 3, the $T^n$ torus can be replaced by any Ricci flat space of the
same dimension.  For instance, K3 can replace $T^4$, and any
Calabi-Yau 6-manifold can replace $T^6$.  These lead to
lower-dimensional theories with less supersymmetry.  Let us consider a
specific example, namely the compactification of type IIA and type IIB
on the K3 manifold.  The resulting six-dimensional theories are also related
by a T-duality in $D=5$, after a compactification on $S^1$.   Both 
six-dimensional theories admit self-dual \cite{lublack}, and more
generally, dyonic \cite{Rahmfeld}),  string solutions
whose horizons have the metric form AdS$_3\times S^3$
\cite{DGT}.  In the case of the type IIB six-dimensional theory, there
exist R-R 3-forms which can support the dyonic string solutions.
Applying the Hopf T-duality on the $U(1)$ fibre coordinate of
$S^3$, we obtain the structure AdS$_3\times CP^1\times S^1$.  If it is
oxidised to $D=10$, we then have a solution of the form AdS$_3\times
CP^1\times S^1\times$K3. If it is further oxidised to $D=11$, it
becomes AdS$_3\times CP^1\times T^2\times$K3.  

\section{Acknowledgements}

We have enjoyed useful conversations with Karim Benakli, Eugene Cremmer, 
Juan Maldacena and Ergin Sezgin.

\vfill\eject

\appendix
\section{T-duality of type IIA and type IIB}

        The Lagrangian of $D=9$, $N=2$ supergravity as the low-energy
limit of type IIA string compactified on a circle can be obtained from
the dimensional reduction of type IIA supergravity in $D=10$, which
itself can be obtained from dimensional reduction of
eleven-dimensional supergravity.  Using the notation adopted in
\cite{lpsol}, the bosonic sector of the theory contains the vielbein,
a dilaton $\phi$ together with a second dilatonic scalar $\varphi$,
(which measures the size of the compactifying circle,) one 4-form
field strength $\td F_4 =dA_3$, two 3-forms $\td F_3^{(i)}=
dA_2^{(i)}$, three 2-forms $\td F_2^{(12)} = dA_1^{(12)}$ and $\td
{\cal F}_2^{(i)}= d {\cal A}_1^{(i)}$ and one 1-form $\td {\cal
F}_1^{(12)} = d{\cal A}_{0}^{(12)}$.  The full bosonic Lagrangian is
given by \cite{lpsol,lpsweyl}
\bea
e^{-1} {\cal L}_{\rm IIA} &=& R - \ft12 (\del \phi)^2 -
\ft12(\del\varphi)^2 - \ft12 ({\cal F}_1^{(12)})^2
e^{\ft32\phi +\ft{\sqrt7}{2} \varphi} \nonumber\\
&& -\ft1{48} (F_4)^2 e^{\ft12\phi +\ft3{2\sqrt7} \varphi} 
-\ft1{12} (F_3^{(1)})^2 e^{-\phi +\ft1{\sqrt7} \varphi} 
- \ft1{12} (F_3^{(2)})^2 e^{\ft12\phi - \ft{5}{2\sqrt7} \varphi}
\label{d92alag}\\
&& -\ft14 (F_2^{(12)})^2 e^{-\phi - \ft3{\sqrt7}\varphi} 
-\ft14 ({\cal F}_2^{(1)})^2 e^{\ft32 \phi +\ft1{2\sqrt7}\varphi} 
- \ft14 ({\cal F}_2^{(2)})^2 e^{\ft4{\sqrt7} \varphi}\nonumber\\
&&-\ft1{2e}\, \td F_4 \wedge \td F_4 \wedge A_1^{(12)} -
\ft1{e}\,\td F_3^{(1)} \wedge \td F_3^{(2)} \wedge A_3\ .\nonumber
\eea
Here we are using the notation that field strengths without tildes
include the various Chern-Simon modifications, whilst field strengths
written with tildes are unmodified.  Thus we have
\bea 
F_4&=&\td F_4 - \td F_3^{(1)}\wedge {\cal A}_1^{(1)} -
\td F_3^{(2)}\wedge {\cal A}_1^{(2)} - \ft12 \td F_2^{(12)}
\wedge {\cal A}_{1}^{(1)}
\wedge {\cal A}_1^{(2)}\ ,\nonumber\\ F^{(1)}_3 &=& \td F^{(1)}_3 - \td
F_2^{(12)} \wedge {\cal A}_1^{(2)}\ ,
\nonumber\\ 
F_3^{(2)} &=& \td F_3^{(2)} + F_2^{(12)}\wedge {\cal
A}_1^{(1)} -
{\cal A}_0^{(12)} (\td F^{(1)}_3 -F_2^{(12)}\wedge {\cal A}_1^{(2)})
\ ,\label{cs9d}\label{csterms}\\ 
F_2^{(12)} &=& \td F^{(12)}_2\ ,\qquad {\cal
F}_2^{(1)} = {\cal F}_2^{(1)} +{\cal A}_0^{(12)} {\cal F}_1^{(2)}
\ ,\quad {\cal F}_2^{(2)} = \td {\cal F}_2^{(2)}\ , 
\quad {\cal F}_1^{(12)} = \td {\cal F}_1^{(12)}\ .\nonumber
\eea

            The Lagrangian of the nine-dimensional supergravity as the
low-energy limit of the type IIB string compactified on a circle can
be obtained from the dimensional reduction of type IIB supergravity in
$D=10$. It is given by
\bea
e^{-1} {\cal L}_{\rm IIB} &=& R-\ft12 (\del\phi)^2 -\ft12 (\del
\varphi)^2 - \ft12 e^{2\phi} (\del \chi)^2 \nonumber\\
&&-\ft1{48} e^{-\ft2{\sqrt7}
\varphi} F_4^2 -\ft1{12} e^{-\phi+\ft1{\sqrt7}\varphi} (F_3^{({\rm
NS})})^2 -\ft12 e^{\phi +\ft1{\sqrt7} \varphi} (F_3^{({\rm R})})^2
\label{d92blag}\\
&&-\ft14 e^{\ft4{\sqrt7} \varphi} ({\cal F}_2)^2 -
\ft14 e^{\phi - \ft3{\sqrt7}\varphi} (F_2^{({\rm R})})^2 -
\ft14 e^{-\phi - \ft3{\sqrt7} \varphi} (F_2^{({\rm NS})})^2\nonumber\\
&&-\ft1{2e} \,\td F_4 \wedge \td F_4 \wedge {\cal A}_1 -
\ft1{e}\, \td F_3^{({\rm NS})} \wedge \td F_3^{({\rm R})} \wedge A_3\ .
\nonumber 
\eea
Note that in $D=10$ there are two 2-form potentials, one of which is
the NS-NS field $A_2^{({\rm NS})}$, and the other is the R-R field
$A_2^{({\rm R})}$.  The dimensional reduction of these two potentials
gives rise to two 2-form potentials and also two vector potentials in
$D=9$, denoted by $A_1^{({\rm NS})}$ and $A_1^{({\rm R})}$
respectively.

        The $D=10$ IIA string and IIB string are related by a
perturbative T-duality, in that type IIA string compactified on a
circle with radius $R$ is equivalent to type IIB string compactified
on a circle with radius $1/R$.  At the level of their low-energy
effective actions, this implies that there is only one $D=9$, $N=2$
supergravity.  The Lagrangians (\ref{d92alag}) and (\ref{d92blag}) are
related to each other by local field redefinitions.  The relations
between the gauge potentials of these two nine-dimensional theories
(including the axions) are summarised in Table 4

\bigskip\bigskip
\begin{center}
\begin{tabular}{|c|c|c|c|c|c|}\hline
    &\multicolumn{2}{|c|}{IIA} &
    &\multicolumn{2}{c|}{IIB} \\ \cline{2-6}
    & $D=10$ & $D=9$ &T-duality & $D=9$ & $D=10$ \\ \hline\hline
    & $A_3$ & $A_3$ & $\longleftrightarrow$ &
                   $A_3$ & $B_4$ \\ \cline{3-6}
R-R & &  $A_2^{(2)}$& $\longleftrightarrow$
                           & $A_2^{\rm R}$ & $A_2^{\rm R}$
                                               \\ \cline{2-5}
fields& ${\cal A}_1^{(1)}$ & ${\cal A}_1^{(1)}$ &
                $\longleftrightarrow$ &
        $A_1^{\rm R}$ & \\ \cline{3-6}
   & & ${\cal A}_0^{(12)}$ & $\longleftrightarrow$
                            & $\chi$ &$\chi$
                                 \\ \hline\hline
NS-NS & $G_{\mu\nu}$ & ${\cal A}_1^{(2)}$
                        & $\longleftrightarrow$ &
        $A_1^{\rm NS}$ & $A_2^{\rm NS}$ \\ \cline{2-5}
fields& $A_2^{(1)}$ & $A_2^{(1)}$ &
               $\longleftrightarrow$ & $A_2^{\rm NS}$ &
                                       \\ \cline{3-6}
      & & $A_1^{(12)}$ & $\longleftrightarrow$ &
                              ${\cal A}_1$ & $G_{\mu\nu}$
                                       \\ \hline
\end{tabular}
\end{center}

\bigskip\bigskip

\centerline{Table 4: Gauge potentials of type II theories in $D=10$
and $D=9$}
\bigskip\bigskip

    The relation between the dilatonic scalars of the two
nine-dimensional theories is given by
\be
\pmatrix{\phi \cr \varphi}_{IIA} =\pmatrix{\ft34 & -\ft{\sqrt7}{4} \cr
                                           -\ft{\sqrt7}{4} & -\ft34}
\pmatrix{\phi \cr \varphi}_{IIB} \ .\label{dils}
\ee
The dimensional reduction of the ten-dimensional string metric to
$D=9$ is given by
\bea
ds_{\rm str}^2 &=& e^{\ft12\phi}\, ds_{10}^2 \nn\\
&=& e^{\ft12\phi}\, (e^{-\varphi/(2\sqrt7)}\, ds_9^2 +
e^{\sqrt7\varphi/2} \, (dz_2 + {\cal A})^2 ) \ ,
\eea
where $ds_{10}^2$ and $ds_9^2$ are the Einstein-frame metrics in
$D=10$ and $D=9$.  The radius of the compactifying circle, measured using
the ten-dimensional string metric, is therefore given by $R=e^{\ft14
\phi +\sqrt7\varphi /4}$.  It follows from (\ref{dils}) that the radii
$R_{IIA}$ and $R_{IIB}$ of the compactifying circles, measured using
their respective ten-dimensional string metrics, are related by
$R_{IIA}=1/R_{IIB}$.

\end{document}